# REGIONAL GRADIENT STRATEGIC SENSORS CHARACTERIZATIONS

**Raheam A. Al-Saphory[1]\*, Naseif J. Al-Jawari[2] and Assma N. Al-Janabi[2]**

[1]Department of Mathematics, College of for the Pure Sciences Education, Tikrit University, Tikrit, IRAQ.
Email: saphory@hotmail.com

[2]Department of Mathematics, College of Science, Al-Mustansiriyah University, Baghdad, IRAQ.

*To whom correspondence should be addressed

## ABSTRACT

In the present paper, the characterizations of regional gradient strategic sensors notions have been given for different cases of regional gradient observability. The results obtained are applied to two dimensional linear infinite distributed system in Hilbert space where the dynamic is governed by strongly continuous semi-group. Various cases of regional strategic sensors are considered and analyzed in connection with regional gradient strategic sensors concepts. Also, we show that there is a various sensors which are not gradient strategic in usual sense for the considered systems, but may be regionally gradient strategic of this system.

**Academic Discipline And Sub-Disciplines:**

Control Systems and Analysis/ Regional strategic sensors structuresfor the systems state gradient.

**TYPE (METHOD/APPROACH)**

Mathematical approach for parabolic distributed parameter systemsgvorned by semi-group operator in state a Hilbert space.

## 1. NTRODUCTION

The analysis of distributed parameter systems refers to a set of concepts such as controllability, observability, detectability [13-14, 18]. The study of these concepts can be made via actuators and sensors structures see [14-17], these concepts give an important link between a system and it's environment [15-18], so that the concepts of actuators and strategic sensors for a class of distributed parameter systems are introduced in order that controllability and observability can be achieved [14-18]. The regional analysis is one of the most important notion of system theory [20-22], it consist to reconstruction the state observation on a sub-region $\omega$ of spatial domain $\Omega$ in finite time[19-23, 25-28], this concepts introduced and developed by El-Jai $et$ $al$. An important extended to the asymptotic case for infinite time by El-Jai and Al-Saphory in several works [1-7]. The study of regional gradient observability for a diffusion system has been given in [27-28] where one is interested in knowledge of the state gradient only in a critical sub-region of the system domain without the knowledge of the state itself. Moreover, the applications are motivated by many real world see [10-12, 21]. Commercial buildings are responsible for a significant fraction of the energy consumption and greenhouse gas emissions in the U.S. and worldwide. Consequently, the design, optimization and control of energy efficient buildings can have a tremendous impact on energy cost and greenhouse gas emission. Mathematically, building models are complex, multi-scale, multi-physics, highly uncertain dynamical systems with wide varieties of disturbances [10].

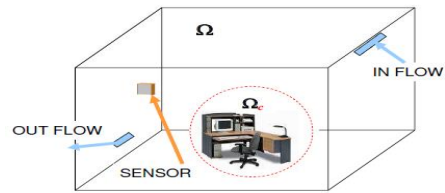

**Fig. 1: Room control model with sensor, in flow and out flow**

In this paper we use a model problem to illustrate that distributed parameter control based on PDEs, combined with high performance computing can be used to provide practical insight into important issues such as optimal sensor/actuator placement (may be best or strategic sensors/ actuators) and optimal supervisory building control. In order to illustrate some of the ideas, we consider the problem illustrated by a single room shown in (Figure 1). This model one can reformulated [11] as spatial case of more general model of distributed parameter systems and represented in the next section (see Figure 2). In addition, the characterization of regional strategic sensors have been given for various types of regional observability in [7].

The purpose of this paper is to extended these results in [7] to the case of regional gradient sensors. Thus, we give a characterization of regional gradient strategic sensors for different cases of regional gradient observation. Therefore, we study and analyze the relationship between the regional gradient strategic sensors and the regional exactly gradient observability. So, the outline of this paper is organized as follows:

Section 2 is present problem statement and basic definitions with characterization of the regional gradient observability. The mathematical concepts of regional gradient strategic sensors in a various situations are studied and developed in section 3. In the last section we gives an application about different sensors locations.

## 2. REIONAL GRADIENT OBSERVABILITY

In this section, we are interested to recall the notion of regional gradient observability and give original results related to particular systems as in [27-28].

### 2.1 Problem Statement

Let $\Omega$ be a regular bounded open subset of $R^n$, with a smooth boundary $\partial\Omega$ and $\omega$ be a non-empty given sub-region of $\Omega$. Let $[0,T], T > 0$ be a time of measurement interval. We denoted $Q = \Omega \times ]0,T[$ and $\Sigma = \partial\Omega \times ]0,T[$. Consider the following distributed parabolic defined by

$$\begin{cases} \frac{\partial x}{\partial t}(\xi,t) = Ax(\xi,t) + Bu(t) & \text{in } Q \\ x(\xi,0) = x_0(\xi) & \text{in } \Omega \\ x(\eta,t) = 0 & \text{in } \Sigma \end{cases} \qquad (1)$$

with the measurements given by the output function

$$y(\cdot,t) = Cx(\,,t) \qquad (2)$$

We have

$$A = \sum_{i,j=1}^{n} \frac{\partial}{\partial x_j}\left(a_{ij}\frac{\partial}{\partial x_j}\right), \text{ with } a_{ij} \in \mathcal{D}(Q).$$

Suppose that $-A$ is elliptic, $i.e.$, there exists $\alpha > 0$ such that

$$\sum_{i,j=1}^{n} a_{ij}\xi_i\xi_j \geq \alpha \sum_{j=1}^{n} |\xi_j|^2, \text{ almost everywhere (a.e) on } Q, \forall \xi = (\xi_1, \dots, \xi_n) \in R^n.$$

This operator is a second order linear differential operator, which generator a strongly continuous semi-group $(S_A(t))_{t\geq 0}$ on the Hilbert space $X = H^1(\Omega)$ and is self-adjoint with compact resolvent. The operator $B \in L(R^p, X)$ and $C \in L(X, R^q)$, depend on the structure of actuators and sensors [18]. The space $X, U$ and $\mathcal{O}$ be separable Hilbert spaces where $X$ is the state space, $U = L^2(0, T, R^p)$ is the control space and $\mathcal{O} = L^2(0, T, R^q)$ is the observation space where $p$ and $q$ are the numbers of actuators and sensors (see Figure 2).

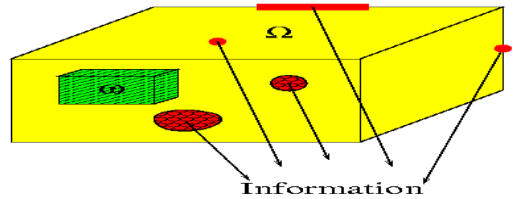

**Fig. 2: The domain of $\Omega$ , the sub-region $\omega$ , various sensors locations**

Under the given assumption, the system (1) has a unique solution [24]:

$$x(\xi, t) = S_A(t)x_0(\xi) + \int_0^t S_A(t - \tau)Bu(\tau)d\tau \tag{3}$$

The problem is to provide sufficient conditions to ensure that, how to extend the results in [7], so that to give a sufficient conditions of regional gradient strategic sensors which achieve the observability in sub-region $\omega$ using various regional gradient sensors.

## 2.2 Definitions And Characterizations

The regional gradient observability concept and reconstruction gradient state have been introduced by Zerrik E. *et al.* as in ref.s [27-28] and recently this concept is developed to the regional asymptotic case by Al-Saphory R [1-7]. Consider the autonomous system to (1) given by

$$\begin{cases} \frac{\partial x}{\partial t}(\xi, t) = Ax(\xi, t) & \text{in } Q \\ x(\xi, 0) = x_0(\xi) & \text{in } \Omega \\ x(\eta, t) = 0 & \text{in } \Sigma \end{cases} \tag{4}$$

The solution of (4) is given by the following form,

$$x(\xi, t) = S_A(t)x_0(\xi), \quad \forall t \in [0, T] \tag{5}$$

The measurements are obtained through the output function

$$y(., t) = Cx(\xi, t)$$

◈ We first recall a sensors is defined by any couple $(D, f)$, where $D$ is spatial support represented by a nonempty part of $\overline{\Omega}$ and $f$ represents the distribution of the sensing measurements on $D$.

Depending on the nature of $D$ and $f$, we could have various type of sensors. A sensor may be pointwise if $D = \{b\}$ with $b \in \overline{\Omega}$ and $f = \delta(.-b)$, where $\delta$ is the Dirac mass concentrated at $b$. In this case the operator $C$ is unbounded and the output function (2) can be written in the form [13-14]

$$y(t) = x(b, t)$$

◈ It may be zonal when $D \subset \overline{\Omega}$ and $f \in L^2(D)$. The output function (2) can be written in the form

$$y(t) = \int_D x(\xi, t)f(\xi)$$

◈ Now , we define the operator

$$K: x \in X \to Kx = CS_A(.)\, x \in \mathcal{O} \tag{6}$$

Thus, we get that

$$y(\cdot, t) = K(t)x(\cdot, 0)$$

where $K$ is bounded linear operator [8].

◈ We note that $K^*: \mathcal{O} \to X$ is the adjoint operator of $K$ defined by

$$K^* y^* = \int_0^t S_A^*(s) C^* y^*(s) ds \qquad (7)$$

◈ Consider the operator

$$\nabla : \begin{cases} H^1(\Omega) & \to (L^2(\Omega))^n \\ x \to \nabla x = (\frac{\partial x}{\partial \xi_1}, \dots, \frac{\partial x}{\partial \xi_n}) \end{cases} \qquad (8)$$

It's adjoint $\nabla^*$ is given by

$$\nabla^* : \begin{cases} (L^2(\Omega))^n \to H^1(\Omega) \\ x \qquad \to \nabla^* x = v \end{cases} \qquad (9)$$

where $v$ is a solution of the Dirichlet problem

$$\begin{cases} \Delta v = -\text{div}(x) & \text{in } \Omega \\ v = 0 & \text{in } \partial\Omega \end{cases}$$

◈ For a nonempty subset $\omega$ of $\Omega$, we consider the operators

$$\chi_\omega : \begin{cases} (L^2(\Omega))^n \to (L^2(\omega))^n \\ x \qquad \to \quad \chi_\omega x = x \mid_\omega \end{cases} \qquad (10)$$

and

$$\tilde{\chi}_\omega : \begin{cases} L^2(\Omega) \to L^2(\omega) \\ x \to \tilde{\chi}_\omega x = x \mid_\omega \end{cases} \qquad (11)$$

where $x \mid_\omega$ is the restriction of $x$ to $\omega$ [9].

◇ Their adjoints are respectively denoted by $\chi_\omega^*$ and $\tilde{\chi}_\omega^*$ are defined by

$$\chi_\omega^* : \begin{cases} (L^2(\omega))^n & \to & (L^2(\Omega))^n \\ x \to \chi_\omega^* x = \begin{cases} x \mid_\omega \text{ in } \omega \\ 0 \quad \text{ in } \Omega \setminus \omega \end{cases} \end{cases} \qquad (12)$$

and

$$\tilde{\chi}_\omega^* : \begin{cases} L^2(\omega) \to L^2(\Omega) \\ x \to \tilde{\chi}_\omega^* x = \begin{cases} x \mid_\omega \quad \text{ in } \omega \\ 0 \quad \text{ in } \Omega \setminus \omega \end{cases} \end{cases} \qquad (13)$$

◈ The idea of gradient observability is based on the existence of an operator $H : \mathcal{O} \to (L^2(\omega))^n$ such that $Hy = \nabla x_0$. This is a natural extension of the observability concept [8]. Then we defined the operator $H = \chi_\omega \nabla K^*$ from $\mathcal{O}$ into $(L^2(\omega))^n$ as in [27]. Now, let us denoted the system (4) together with the output (2) by (4)-(2).

**Definition 2.1:** The system (4)-(2) is said to be regionally exactly observable on a sub-region $\omega$ (*exactly $\omega$ - observable*), if

$$Im \tilde{\chi}_\omega K^* = L^2(\omega)$$

**Definition 2.2:** The system (4)-(2) is said to be regionally weakly observable on $\omega$ (*weakly $\omega$- observable*), if

$$\overline{Im \tilde{\chi}_\omega K^*(.)} = L^2(\omega)$$

**Definition 2.3:** The system (4)-(2) is said to be regionally exactly gradient observable on $\omega$ (*exactly $\omega_G$ - observable*), if

$$Im \chi_\omega \nabla K^* = (L^2(\omega))^n$$

**Definition 2.4:** The system (4)-(2) is said to be regionally weakly gradient observable on $\omega$ (*weakly $\omega_G$ - observable*), if

$$\overline{Im \chi_\omega \nabla K^*(.)} = (L^2(\omega))^n$$

We see that if a system is weakly $\omega_G-$ observable then there is one to one relationship between the output and the initial gradient, viz., if $y$ is given and $x_0$ satisfies $y = CS(.)x_0$ then $\nabla x_0$ is unique.

**Remark 2.5:** We can deduced that, the definition 2.4 is equivalent to say that the system (4)-(2) is weakly$\omega_G$-observable, if

$$ker \, K(t)\nabla^* \chi_\omega^* = \{0\}$$

Then, the following characterization can extend to the regional gradient case as in ref.[26]

**Proposition 2.6:** The system (4)-(2) is exactly $\omega_G$-observable if and only if there exist $c > 0$ such that for all $x^* \in (L^2(\omega))^n$, such that,

$$\| \chi_\omega x^* \|_{(L^2(\omega))^n} \le c\|K(t)\nabla^* \chi_\omega^* x^* \|_\mathcal{O} \qquad (14)$$

**Proof:** The proof of this property is deduced from the usual results on observability considering $\chi_\omega \nabla K^*$ [13]. Let $V, W$ and $X$ be a reflexive Banach space and let $F \in L(V, X), G \in L(W, X)$, then the following conditions are equivalent

1. $Im\ F \subset Im\ G$.
2. There exist $c > 0$ such that $\|F^* x^*\|_{V^*} \le c\|G^* x^*\|_{W^*} \qquad \forall x^* \in G^*$

Now, by applying the above result we obtain the equivalent condition for exactly $\omega_G$-observable as:

Let $V = X = (L^2(\omega))^n, W = \mathcal{O}, F = Id_{(L^2(\omega))^n}$ and $G = \chi_\omega \nabla K^*$.

Now, since the system is exactly $\omega_G$-observable we have $Im\ F \subset Im\ G$, which is equivalent to that fact there exist $c > 0$, such that

$$\|F^* x^*\|_{l^2(\omega))^n} \le c\|G^* x^*\|_{W^*} \qquad \forall x^* \in G^*. \square$$

**Remark 2.7:** We have:

(1) The regional state reconstruction will be more precise than the whole domain if we estimate the state in the whole the domain.

(2) From (14) there exists a reconstruction error operator that gives the estimation $\tilde{x}_0$ of the initial state $x_0$ in $\omega$, and then, If we put e $= x_0 - \tilde{x}_0$, we have

$$\|e\|_{(L^2(\omega))^n} \le \|e\|_{(L^2(\Omega))^n}$$

$$\Rightarrow \| x_0 - \tilde{x}_0 \|_{(L^2(\omega))^n} \le \| x_0 - \tilde{x}_0 \|_{(L^2(\Omega))^n}$$

Where, $x_0$ is the exact state of the system and $\tilde{x}_0$ is the estimated state of the system.

**Proposition 2.8:** If the system is exactly $\omega$-observable then it is exactly $\omega_G$-observable.

**Proof:** Since the system is exactly $\omega$-observable there exist $\gamma > 0$ such that $\forall x_0 \in L^2(\omega)$, we have

$$\|x_0\|_{L^2(\omega)} \le \gamma\|K\chi_\omega^* x_0\|_{L^2(0,T,\mathcal{O})} \qquad, \ \forall \gamma > 0$$
since $(L^2(\omega))^n \subset L^2(\omega)$, then

$$\|\nabla x_0\|_{(L^2(\Omega))^n} = \|x_0\|_{(L^2(\omega))^n} \le \|x_0\|_{L^2(\omega)}, \forall\ x_0 \in L^2(\omega), \text{ where}$$

$$L^2(\omega) = \{x_0 : \int_\omega |x_0|^2 < \infty\}, \qquad (15)$$

and then

$$(L^2(\omega))^n = \left\{\nabla x_0 = g_i : \int_\omega |g_i|^2 < \infty,\ g_i = \frac{\partial x_0}{\partial \xi_i} \ \forall i = 1,2,\dots \right\}.$$

To prove $\|x_0\|_{(L^2(\omega))^n} \le c\|K\nabla^* \chi_\omega^* x_0\|_{L^2(0,T,\mathcal{O})}$ then from (15) and since a system exactly $\omega$-observable, then there exist $\gamma > 0$ and $c > 0$ such that $\gamma = \frac{1}{c}$ and by choosing

$$c = \frac{\|K\chi_\omega^* x_0\|_\mathcal{O}}{\|K\nabla^* \chi_\omega^* x_0\|_\mathcal{O}} \qquad (16)$$

and then

$$\|x_0\|_{(L^2(\omega))^n} \le \|x_0\|_{L^2} \le \gamma\|K\tilde{\chi}_\omega^* x_0\|_\mathcal{O} \qquad (17)$$

substitute (16) in (17), we obtain

$$\|x_0\|_{(L^2(\omega))^n} \le \|K\nabla^* \chi_\omega^* x_0\|_\mathcal{O}$$

Therefore this system is exactly $\omega_G$-observable with $\gamma = 1.\square$

**Remark 2.9:** From the above proposition we can get the following result:

If the system is exactly $\omega$-observable then it is exactly G-observable in $\omega^1$ for all $\omega^1 \subset \omega$ (*exactly $\omega_G^1$-observable*).

## 3. REGIONAL GRADIENT STRATEGIC SENSORS

The purpose of this section is to give the characterization for sensors in order that the system (4)-(2) which is observable in $\omega$.

### 3.1 $\omega_G$-Strategic Sensors

**Definition 3.1:** A sensor $(D, f)$ is regional gradient strategic on $\omega$ (*$\omega_G$-strategic*) if the observed system is weakly $\omega_G$-observable.

**Definition 3.2:** A sensor $(D_i, f_i)_{1 \leq i \leq q}$ is $\omega_G$-strategic if there exist at least one sensor $(D_1, f_1)$ which $\omega_G$-strategic.

We can deduce that the following result:

**Corollary 3.3:** A sensor is $\omega_G$-strategic if the observed system is exactly $\omega_G$-observable.

**Proof:** Let the system exactly $\omega_G$-observable, then , we have

$$Im \; \chi_\omega \nabla K^* = (L^2(\omega))^n$$

From decomposition sub-space of direct sum in Hilbert space, we represent $(L^2(\Omega))^n$ by the unique form [13]

$$ker \; \chi_\omega + Im \chi_\omega^* \chi_\omega \nabla K^* = (L^2(\Omega))^n$$

We obtain

$$ker \; K(t) \nabla^* \chi_\omega^* = \{0\}$$

This is equivalent to [9]

$$\overline{Im \; \chi_\omega \nabla K^*(.)} = (L^2(\omega))^n$$

Finally, we can deduce this system is weakly $\omega_G$-observable and therefore this sensor is $\omega_G$-strategic.□

**Corollary 3.4:** A sensor is $\omega_G$-strategic if and only if the operator $N_\omega = HH^*$ is positive definite.

**Proof:** Since a sensor is $\omega_G$-strategic this mean that the system is weakly $\omega_G$-observable,

let $x^* \in (L^2(\omega))^n$ such that

$$< N_\omega x^*, x^* >_{(L^2(\omega))^n} = 0 \quad \text{then} \quad \|H^* x^*\|_O = 0$$

and therefore $x^* \in ker \; H^*$, thus, $x^* = 0$, i.e., $N_\omega$ is positive definite.

Conversely, let $x^* \in (L^2(\omega))^n$ such that

$$H^* x^* = 0, \text{ then } < H^* x^*, H^* x^* >_O = 0$$

and thus,

$$< N_\omega x^*, x^* >_{(L^2(\omega))^n} = 0$$

Hence $x^* = 0$ thus the system is weakly $\omega_G$-observable and therefore a sensor is $\omega_G$-strategic.□

**Remark 3.5:** From the previous results, we obtain that:

(1) If the system is exactly $\omega_G$-observable then the system is weakly $\omega_G$-observable and therefore this sensor is $\omega_G$-strategic.

(2) A sensor which is regional gradient strategic sensor in $\omega^1$ ($\omega_G^1$-*strategic*) for a system where $\omega^1 \subset \Omega$, is regional gradient strategic sensor in $\omega^2$ ($\omega_G^2$-strategic) for any $\omega^2 \subset \omega^1$.

(3) The concept of exact $\omega_G$-observability is more restrictive than weak $\omega_G$-observability.

Now, assume that the operator $A$ has a complete set of eigenfunction in $H^1(\Omega)$, denoted by $\varphi_{nj}$, which is orthonormal in $L^2(\omega)$ and the associated with the eigenvalue $\lambda_n$ of multiplicities $r_n$, then the concept of regional gradient strategic on $\omega$ can be characterized by the following result:

**Theorem 3.6:** Assume that $\sup r_n = r < \infty$, then the suite of sensors $(D_{\phi} f_i)_{1 \leq i \leq q}$, $\omega_G$-strategic if

    (1) $q \geq r$
    (2) $rank \; G_n = r_n \quad \forall n \geq 1, where$

Where $G_n = (G_n)_{ij}$ for $1 \le i \le q, 1 \le j \le r_n$, and

$$(G_n)_{ij} = \begin{cases} \sum_{k=1}^m \frac{\partial \varphi_{nj}}{\partial \xi_k}(b_i) & \text{in the pointwise case} \\ \sum_{k=1}^n < \frac{\partial \varphi_{nj}}{\partial \xi_k}, f_i >_{L^2(D_i)} & \text{in the zonal case} \end{cases}$$

**Proof:** We will discussed the case where the sensors are of pointwise type and located inside the domain $\Omega$. The suite of sensors $(b_i, \delta_{b_i})_{1 \le i \le q}$ is $\omega_G$-strategic if and only if

$$\{x^* \in (L^2(\omega))^n| < Hy, x^* >_{(L^2(\omega))^n} = 0, \forall y \in \mathcal{O}\} \implies x^* = 0.$$

Suppose that the suite of sensors $(b_i, \delta_{b_i})_{1 \le i \le q}$ is $\omega_G$-strategicbut for a certain $n \in N, rank \ G_n \ne r_n$, then there exists a vector $x_n = (x_{n_1}, x_{n_2}, ..., x_{n_{r_n}})^{tr} \ne 0$, such that $G_n x_n = 0$. So, we can construct a nonzero $x_0 \in L^2(\omega)$ considering $< x_0, \varphi_{pj} >_{L^2(\omega)} = 0$ if $p \ne n$ and

$$< x_0, \varphi_{nj} >_{L^2(\omega)} = x_{nj}, 1 \le j \le r_n.$$

Let $x_0 = \sum_{j=1}^{r_n} x_{nj} \varphi_{nj}$ , $x_0 = (x_0, x_0, ..., x_0)$, then

$$< Hy, x_0 >_{(L^2(\omega))^n} = \sum_{k=1}^n < \chi_\omega \frac{\partial}{\partial \xi_k}(K^*y), x_0 >_{L^2(\omega)}$$

$$= \sum_{k=1}^n < \frac{\partial}{\partial \xi_k}(\tilde{x}(T)), \chi_\omega^* x_0 >_{L^2(\Omega)}$$

where $\tilde{x}$ is the solution of the following system:

$$\begin{cases} \frac{\partial \tilde{x}}{\partial t}(\xi, t) = A^* \tilde{x}(\xi, t) + \sum_{i=1}^q \delta_{b_i} y_i(T - t) & \text{in } Q \\ \tilde{x}(\xi, 0) = 0 & \text{in } \Omega \\ \tilde{x}(\eta, t) = 0 & \text{in } \Sigma \end{cases} \qquad (18)$$

Consider the system:

$$\begin{cases} \frac{\partial \varphi}{\partial t}(\xi, t) = -A \varphi(\xi, t) & \text{in } Q \\ \varphi(\xi, 0) = \chi_\omega^* x_0 & \text{in } \Omega \\ \varphi(\eta, t) = 0 & \text{in } \Sigma \end{cases} \qquad (19)$$

Multiply (18) by $\frac{\partial \varphi}{\partial \xi_k}$ and integrate on $Q$, we obtain

$$\int_Q \frac{\partial \varphi}{\partial \xi_k}(\xi, t) \frac{\partial \tilde{x}}{\partial t}(\xi, t) d\xi \ dt = \int_Q A^* \tilde{x}(\xi, t) \frac{\partial \varphi}{\partial \xi_k}(\xi, t) d\xi \ dt \ +$$

$$\int_Q (\sum_{i=1}^q \delta_{b_i} y_i(T - t)) \frac{\partial \varphi}{\partial \xi_k}(\xi, t) d\xi dt.$$

But we have

$$\int_Q \frac{\partial \varphi}{\partial \xi_k}(\xi, t) \frac{\partial \tilde{x}}{\partial t}(\xi, t) d\xi \ dt = \int_\Omega \left[ \frac{\partial \varphi}{\partial \xi_k}(\xi, t) \tilde{x}(\xi, t) d\xi \right]_0^T + \int_Q A \frac{\partial \varphi}{\partial \xi_k}(\xi, t) \tilde{x}(\xi, t) d\xi \ dt$$

$$= \int_\Omega \frac{\partial \varphi}{\partial \xi_k}(\xi, t) \tilde{x}(\xi, t) d\xi + \int_Q A \frac{\partial \varphi}{\partial \xi_k}(\xi, t) \tilde{x}(\xi, t) d\xi \ dt$$

then:

$$\int_\Omega \frac{\partial \varphi}{\partial \xi_k}(\xi, t) \tilde{x}(\xi, t) d\xi = -\int_Q A \frac{\partial \varphi}{\partial \xi_k}(\xi, t) \tilde{x}(\xi, t) d\xi + \int_Q A^* \tilde{x}(\xi, t) \frac{\partial \varphi}{\partial \xi_k}(\xi, t) d\xi \ dt$$

$$+ \int_Q (\sum_{i=1}^q \delta_{b_i} y_i(T - t)) \frac{\partial \varphi}{\xi_k}(\xi, t) d\xi dt$$

integrating by parts we obtain

$$\int_\Omega \frac{\partial \varphi}{\partial \xi_k}(\xi, t) \tilde{x}(\xi, t) d\xi = -\int_\pi \frac{\partial \tilde{x}(\eta, t)}{\partial v_{A^*}} \frac{\partial \varphi}{\partial \xi_k}(\eta, t) d\eta dt$$

$$\int_\pi \frac{\partial}{\partial v_{A^*}}\left(\frac{\partial \varphi}{\partial \xi_k}(\eta,t)\,d\eta dt\right)\tilde{x}(\eta,t)\,d\eta dt + \int_Q \left(\sum_{i=1}^q \delta_{b_i} y_i(T-t)\right)\frac{\partial \varphi}{\partial \xi_k}(\xi,t)\,d\xi dt$$

the boundary conditions give

$$\int_\Omega \frac{\partial \varphi}{\partial \xi_k}(\xi,t)\tilde{x}(\xi,t)\,d\xi = \int_Q \left(\sum_{i=1}^q \delta_{b_i} y_i(T-t)\right)\frac{\partial \varphi}{\partial \xi_k}(\xi,t)\,d\xi dt.$$

Thus

$$\int_\Omega \varphi(\xi,t)\frac{\partial \tilde{x}}{\partial \xi_k}(\xi,T)\,d\xi = -\sum_{i=1}^q \int_0^T \frac{\partial \varphi}{\partial \xi_k}(b_i,t)\,y_i(T-t)\,dt$$

and we have

$$< \chi_\omega \nabla K^* y, x_0 >_{(L^2(\omega))^n} = \sum_{k=1}^n \int_\Omega \frac{\partial \tilde{x}}{\partial \xi_k}(\xi,t)\varphi(\xi,t)\,d\xi$$

$$= -\sum_{K=1}^q \int_0^T \sum_{k=1}^n \frac{\partial \varphi}{\partial \xi_k}(b_i,t)\,y_i(T-t)\,dt.$$

But

$$\varphi(\xi,t) = \sum_{p=1}^\infty e^{-\lambda_p(T-t)}\sum_{j=1}^{r_p} < x_\circ,\varphi_{pj}>_{L^2(\omega)}\varphi_{pj},$$

Then

$$\sum_{k=1}^n \frac{\partial \varphi}{\partial \xi_k}(b_i,t) = \sum_{p=1}^\infty e^{-\lambda_p(T-t)}\sum_{j=1}^{r_p} < x_0,\varphi_{pj}>_{L^2(\omega)}\sum_{k=1}^n \frac{\partial \varphi}{\partial \xi_k}(b_i)$$

$$= \sum_{p=1}^\infty e^{\lambda_p(T-t)}(G_p x_p)_i$$

therefore

$$< \chi_\omega \nabla K^* y, x_0 >_{(L^2(\omega))^n} = -\sum_{i=1}^q \int_0^T \sum_{p=1}^\infty e^{\lambda_p(T-t)}(G_p x_p)_i\,y_i(T-t)\,dt \qquad (20)$$

Thus

$$< \chi_\omega \nabla K^* y, x_0 >_{(L^2(\omega))^n} = -\sum_{i=1}^q \int_0^T e^{\lambda_n(T-t)}(G_n x_n)_i\,y_i(T-t)\,dt = 0$$

This is true for all $y \in L^2(0,T;R^q)$, then $x_0 \in Ker\,H^*$ which contradicts the assumption that the suite of sensors is $\omega_G$-strategic.□

We can deduced the following result:

**Corollary 3.7:** In the one dimension case, a sensor is $\omega_G$-strategic if and only if $q \geq r = sup\,r_n$ and $rank\,G_n = r_n, \forall n \geq 1$, where $G_n$ is given in theorem 3.6.

**Remark 3.8:** From the previous results, we can get

(1) The Theorem 3.6 implies that the required number of sensors is greater than or equal to the largest multiplicity of the eigenvalues.

(2) By infinitesimally deforming the domain, the multiplicity can be reduced to one [19]. Consequently, $\omega_G$-strategic sensors can be achieved using only one sensor.

Now, we can deduced that various sensors which are not strategic in usual sense for systems, but may be $\omega_G$-strategic and achieve the $\omega_G$-observability. This is illustrated in the following counter- example.

### 3.2 A Counter- Example

Consider the system described by the parabolic equation

$$\begin{cases} \frac{\partial x(\xi,t)}{\partial t} = \frac{\partial^2 x}{\partial \xi^2}(\xi,t) & \text{in }]0,1[\times]0,T[ \\ x(0,1) = x(1,t) = 0 & \text{in }]0,T[ \\ x(\xi,0) = x_0(\xi) & \text{in }]0,1[ \end{cases} \qquad (21)$$

Suppose that the measurement is given by pointwise sensor located in $b \in ]0,1[$ which is given by the following output function

$$y(.,t) = \int_\Omega x(\xi,t)\,\delta(\xi-t)\,d\xi = x(b,t),\ t \in (0,T) \qquad (22)$$

Where $\varphi_n = \sqrt{2}\sin(n\pi\xi)$ and $\lambda_n = -n^2\pi^2$. First, we must prove that the system (21)-(22) is not weakly observable in $\Omega$, that means the sensors $(\delta_b, b)$ is not strategic. For this purpose, we can write the system (21) as a state space one dimensional system

$$\dot{x}(\xi, t) = Ax(\xi, t)$$

$$x(\xi, 0) = x_0(\xi)$$

Where $A = \frac{\partial^2}{\partial \xi^2}$ generate the continuous semigroup $(S(t))_{t\geq 0}$ given by [17].

$$S(t)x_0 = \sum_{i=1}^{\infty} e^{\lambda_i t} <x_0, \varphi_i>_{L^2(\Omega)} \varphi_i$$

Where, $\varphi_n = \sqrt{2}\sin(n\pi\xi)$, $\lambda_n = -n^2\pi^2$ are the eigenfunctions associated with the eigenvalues of $A$. Then from solution of (21), we have

$$y(\xi, t) = \sum_{i=1}^{\infty} e^{\lambda_i t} <x_0, \varphi_i>_{L^2(\Omega)} \varphi_i(b) = CS(t)x_0 = K(t)x_0$$

The system (21)-(22) is weakly observable if $\ker K(t) = \{0\}$.

As proved in [27], if $b \in Q$ then system (21)-(22) is not weakly observable on $\Omega=(0,1)$ and a sensor $(\delta_b, b)$ is not strategic.

A sensor is $\omega$-strategic on $(0,1)$ $\Leftrightarrow$ $:b \in S = \bigcup_{n=1}^{\infty} \left\{\frac{k}{n} | k \in [1, n-1] \cap N\right\}$. Since $\sin(n\pi b) = 0 \Leftrightarrow nb = k \Rightarrow b = \frac{k}{n}$. Consequently, the system is weakly observable on $(0,1)$. And then, it is G-strategic on $(0,1)$ $\Leftrightarrow$ $b \notin S_G = \bigcup_{n=1}^{\infty} \left\{\frac{2k+1}{2n} | k \in [0, n-1] \cap N\right\}$. Since $\cos(n\pi b) = 0 \Leftrightarrow nb = \frac{2k+1}{2} \Rightarrow b = \frac{2k+1}{2n}$. Consequently, the system is weakly G-observable on $(0,1)$.□

**Corollary 3.9:** If the system (21)-(22) is exactly $\omega_G$-observable, rank condition in theorem (3.6) is satisfied and a sensor is $\omega_G$-strategic.

Now, assume that a sensor is not gradient strategic in whole the domain $\Omega$ and let $(\bar{\varphi}_i)_{i\in N^n}$ be a basis in $(L^2(\Omega))^n$. Let $I \subset N^n$ be such that $\ker KV^* = span \{(\bar{\varphi}_i)_{i\in I}\}$ and $J = N^n \backslash I$.

Now, assume that a sensor is not gradient strategic in whole the domain $\Omega$ and let $(\bar{\varphi}_i)_{i\in N^n}$ be a basis in $(L^2(\Omega))^n$. Let $I \subset N^n$ be such that $\ker KV^* = span \{(\bar{\varphi}_i)_{i\in I}\}$ and $J = N^n \backslash I$.

**Proposition 3.10:** The following properties are equivalent:

1. A sensor is $\omega_G$-strategic.
2. $\overline{span \{(\chi_\omega \bar{\varphi}_i)_{i\in J}\}} = (L^2(\omega))^n$
3. If $x \in (L^2(\omega))^n$ is such that $<x, \chi_\omega \bar{\varphi}_i>_{(L^2(\omega))^n} = 0$ for all $i \in J$, then $x = 0$.
4. If $\sum_{i\in I} a_i \bar{\varphi}_i = 0$ in $\Omega \backslash \omega$, then $a_i = 0$ for all $i \in I$.

**Proof:** 1$\Rightarrow$2 Since sensors are $\omega_G$-strategic this mean that the system is weakly $\omega_G$-observable.

Let $x \in (L^2(\omega))^n$. Then for $\varepsilon > 0$ $\exists$ $y \in \mathcal{O}$ such that

$$\|x - \chi_\omega \nabla K^*\|_{(L^2(\omega))^n} \leq \varepsilon, \text{ but}$$

$$\nabla K^* y = \sum_{i\in N^n} <\nabla K^* y, \bar{\varphi}_i>_{(L^2(\Omega))^n} \bar{\varphi}_i = \sum_{i\in J} <y, K\nabla^*\bar{\varphi}_i>_{\mathcal{O}} \bar{\varphi}_i, \text{ and thus}$$

$$\chi_\omega \nabla K^* y = \sum_{i\in J} <y, K\nabla^*\bar{\varphi}_i>_{\mathcal{O}} \chi_\omega \bar{\varphi}_i. \text{ Then}$$

$$\left\| x - \sum_{i\in J} <y, K\nabla^*\bar{\varphi}_i>_{\mathcal{O}} \chi_\omega \bar{\varphi}_i \right\|_{(L^2(\omega))^n} < \varepsilon$$

and hence $x \in \overline{\{\chi_\omega \bar{\varphi}_i\}_{i\in J}}$.

2$\Rightarrow$3 Let $x \in (L^2(\omega))^n$. For any $\varepsilon > 0$ $\exists$ $\alpha_j (j \in J)$ such that

$$\left\| x - \sum_{i\in J} \alpha_j \chi_\omega \bar{\varphi}_j \right\|^2_{(L^2(\omega))^n} < \varepsilon, \text{ with}$$

$$<x, \chi_\omega \bar{\varphi}_j>_{(L^2(\omega))^n} = 0, \forall j \in J$$

we deduced that

$$\|x\|^2_{(L^2(\omega))^n} < \varepsilon. \text{ Thus, } x = 0.$$

3$\Rightarrow$4 Let $\sum_{i\in I} a_i \bar{\varphi}_i = 0$ in $\Omega \backslash \omega$.

Now Consider $x = \chi_\omega(\sum_{i\in I} a_i \bar{\varphi}_i)$. For $j \in J$, we have

$$< x, \chi_\omega \bar{\varphi}_i >_{(L^2(\omega))^n} = \sum_{i\in I} a_i < \bar{\varphi}_i, \bar{\varphi}_j >_{(L^2(\Omega))^n} = 0.$$

Since $x = 0$, we get

$$\sum_{i\in I} a_i \bar{\varphi}_i = 0 \text{ in } \Omega \text{ and } a_i = 0, \forall i \in I.$$

$4 \Longrightarrow 1$ Consider $x \in (L^2(\omega))^n$ such that

$K \nabla^* \chi_\omega^* x = 0$. We have $\chi_\omega^* x \in (L^2(\Omega))^n$

then

$K \nabla^* \chi_\omega^* x = K \nabla^*(\sum_{i\in N^n} < x, \chi_\omega \bar{\varphi}_i >_{(L^2(\omega))^n} \bar{\varphi}_i) = K \nabla^*(\sum_{i\in j} < x, \chi_\omega \bar{\varphi}_i >_{(L^2(\omega))^n} \bar{\varphi}_i) = 0$. Therefore,

$$\sum_{i\in j} < x, \chi_\omega \bar{\varphi}_i >_{(L^2(\omega))^n} \bar{\varphi}_i \in span\{(\bar{\varphi}_i)_{i\in I}\}$$

and then

$$\sum_{i\in I} < x, \chi_\omega \bar{\varphi}_i >_{(L^2(\omega))^n} = 0, \forall j \in J.$$

Therefore

$$\chi_\omega^* x = \sum_{i\in I} < x, \chi_\omega \bar{\varphi}_i >_{(L^2(\omega))^n} \bar{\varphi}_i = 0 \text{ in } \Omega\backslash\omega.$$

From the assumption we have $< x, \chi_\omega \bar{\varphi}_i >_{(L^2(\omega))^n} = 0, \forall i \in I$. Hence $x = 0.\square$

We can deduced the following result:

**Corollary 3.11:** Under the hypotheses of Proposition 3.10, a sensors is $\omega_G$-strategic in all $\omega \subset \Omega$ such that $< \bar{\varphi}_i, \bar{\varphi}_j >_{(L^2(\omega))^n} = 0, \ \forall i, j \in I, i \neq j$.

**Proof:** To deduce the result from previous Proposition 3.10, we take $\sum_{i\in I} a_i \bar{\varphi}_i = 0$ in $\Omega\backslash\omega$. Then we only need to show that $a_i = 0, \forall i \in I$. Let $x = \sum_{i\in I} a_i \bar{\varphi}_i$ in $\bar{\Omega}$ and $i_0 \in I$. Then

$$< x, \bar{\varphi}_{i0} >_{(L^2(\Omega))^n} = < \sum_{i\in I} a_i \bar{\varphi}_i, \bar{\varphi}_{i0} >_{(L^2(\Omega))^n} = \sum_{i\in I} a_i < \bar{\varphi}_i, \bar{\varphi}_{i0} >_{(L^2(\Omega))^n} = a_{i0} \qquad (23)$$

Since $x = 0$ in $\Omega\backslash\omega$, under the assumption of Corollary 3.11 we have

$$< x, \bar{\varphi}_{i0} >_{(L^2(\Omega))^n} = \sum_{i\in I} a_i < \bar{\varphi}_i, \bar{\varphi}_{i0} >_{(L^2(\omega))^n} = a_{i0} \|\bar{\varphi}_{i0}\|^2_{(L^2(\omega))^n} \qquad (24)$$

From (23)-(24), we obtain $a_i = 0, \forall i \in I$.

## 4. APPLICATION TO SENSORS LOCATIONS

In this section, we give specific results related to the different case presented in the above section. First we consider internal sensors (zonal, pointwise, filament in rectangular and disk domain) the presented result give information on the structure of $\omega$. Consider the system

$$\begin{cases} \frac{\partial x}{\partial t}(\xi_1, \xi_2, t) = \Delta x(\xi_1, \xi_2, t) & \text{in } Q, \\ x(\xi_1, \xi_2, 0) = x_0(\xi_1, \xi_2) & \text{in } \Omega, \\ x(\eta_1, \eta_2, t) = 0 & \text{in } \Sigma \end{cases} \qquad (25)$$

Let $\Omega = (0,1) \times (0,1)$ and let $\omega = (\alpha_1, \beta_1) \times (\alpha_2, \beta_2)$ be the considered region is subset of $\Omega$, the eigenfunctions and the eigenvalue of the system (25) are given by:

$$\varphi_{ij}(\xi_1, \xi_2) = \frac{2}{\sqrt{(\beta_1-\alpha_1)(\beta_2-\alpha_2)}} \sin i\pi \frac{(\xi_1-\alpha_1)}{(\beta_1-\alpha_1)} sin j\pi \frac{(\xi_2-\alpha_2)}{(\beta_2-\alpha_2)} \qquad (26)$$

Associated with eigenvalue

$$\lambda_{ij} = -\frac{i^2}{(\beta_1-\alpha_1)^2} + \frac{j^2}{(\beta_2-\alpha_2)^2} \qquad (27)$$

### 4.1 Internal Zone Sensor

Consider the system (25) together with output function (2) where the sensor supports $D$ are located in $\Omega$. The output (2) can by written by the form

$$y(t) = \int_D x(\xi_1, \xi_2, t) f(\xi_1, \xi_2) d\xi_1 d\xi_2 \qquad (28)$$

Where $D \subset \Omega$ is location of zone sensor and $f \in L^2(D)$. In this case of (see Figure 3), the eigenfunctions and the eigenvalues (26) and (27).

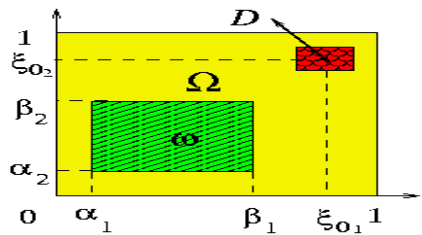

**Fig. 3: Domain $\Omega$, sub-region $\omega$ and location $D$ of internal zone sensor**

However , if we suppose that

$$\frac{(\beta_1 - \alpha_1)^2}{(\beta_2 - \alpha_2)^2} \notin Q$$

Then multiplicity of $\lambda_{ij}$ is $r_{ij} = 1$ and then one sensor $(D, f)$ my be sufficient to achieve $\omega_G$-observable of the systems (25) and (28) [19].Let the measurement support is rectangular with

$$D = [\xi_{01} - l_1, \xi_{01} + l_1] \times [\xi_{02} - l_2, \xi_{02} + l_2] \in \Omega$$

Then, we have the following result

**Corollary 4.1:** If $f_1$ is symmetric about $\xi_1 = \xi_{01}$ and $f_2$ is symmetric about $\xi_2 = \xi_{02}$, then the sensor $(D, f)$ is $\omega_G$-strategic if

$$\frac{i(\xi_{01} - \alpha_1)}{(\beta_1 - \alpha_1)} \text{ and } \frac{j(\xi_{02} - \alpha_1)}{(\beta_2 - \alpha_2)} \notin N \text{ for some } i, j.$$

### 4.2 Internal Pointwise Sensor

In this case the out put function is given by:

$$y(t) = \int_D x(\xi_1, \xi_2, t) \delta(\xi_1 - b_1, \xi_2 - b_2) d\xi_1 d\xi_2 \qquad (29)$$

With $b = (b_1, b_2)$ is location of pointwise sensor as defined in (see Figure 4)

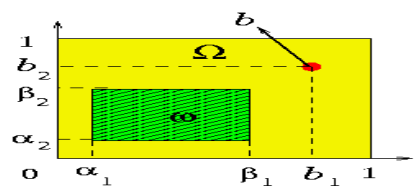

**Fig. 4: Rectangular domain, and location $b$ of internal pointwise sensor**

If $\frac{(\beta_1 - \alpha_1)}{(\beta_2 - \alpha_2)} \notin Q$, then $r_{ij} = 1$ and one sensor $(b, \delta_b)$ may be sufficient for $\omega_G$-observability of the systems (25)-(29)

**Corollary 4.2:** The sensor $(b, \delta_b)$ is $\omega_G$-strategic if

$$\frac{i(b_1 - \alpha_1)}{(\beta_1 - \alpha_1)} \text{ and } \frac{j(b_2 - \alpha_2)}{(\beta_2 - \alpha_2)} \notin N, \text{ for some } i, j.$$

### 4.3 Internal Filament Sensor

Consider the case where the observation is given on the curve $\sigma = Im(\gamma)$ with $\gamma \in C^1(0,1)$ (see Figure 5)

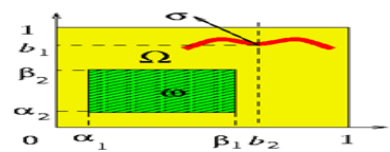

**Fig. 5: Rectangular domain, and location $\sigma$ of internal filament sensors**

**Corollary 4.3:** If the measurements recovered by filament sensor $(\sigma, \delta_\sigma)$ such that is symmetric with respect to the line $\xi = \xi_0$. Then the sensor $(\sigma, \delta_\sigma)$ is $\omega_G$-strategic if

$$\frac{i(\xi_{01} - \alpha_1)}{(\beta_1 - \alpha_1)} \text{ and } \frac{j(\xi_{02} - \alpha_2)}{(\beta_2 - \alpha_2)} \notin N, \text{ for } i, j = 1, \dots, J.$$

**Remark 4.4:** These results can be extended to the following:

1. Case of Neumann or mixed boundary conditions [4-5].

2. Case of disc domain $\Omega = (D, 1)$ and $\omega = (0, r_\omega)$ where $\omega \subset \Omega$ and $0 < r_\omega < 1$ [1-3].

3. Case of boundary sensors where $C \notin L(X, R^q)$, we refer to see [13-14].

4. We can show that the observation error decreases when the number and support of sensors increases [23, 25].

## 5. CONCLUSION

We have been introduced a sufficient condition of regional gradient strategic sensors in order to achieves regional gradient observability. Many interesting results concerning the choice of sensors structure are given and illustrated in specific situations. Various questions still opened under consideration. For example, these result can be extended to the boundary case with parabolic and hyperbolic systems [8].

### ACKNOWLEDGMENTS

Our thanks in advance to the editors and experts for considering this paper to publish in this estimated journal and for their efforts.

# REFERENCES

[1] Al-Saphory, R. and El Jai, A. 2001. Sensor structures and regional detectability of parabolic distributes systems. International Journal of Sensors and Actuators. Vol. 90.163-171.

[2] Al-Saphory, R. and El Jai, A. 2001. Sensors characterizations for regional boundary detectability in distributed parameter systems. International Journal of Sensors and Actuators. Vol. 94.1-10.

[3] Al-Saphory, R. and El Jai A. 2002. Regional asymptotic state reconstruction. International Journal of Systems Science. Vol. 33. 1025-1037.

[4] Al-Saphory, R. 2010. Strategic sensors and regional exponential observation in neumann boundary conditions. Zanco Journal of Pure and Applied Science. Vol. 22, 1-11.

[5] Al-Saphory, R., Al-Jawari, N. and Al-Qaisi, A . 2010. Regional gradient detectability for infinite dimensional systems. Tikrit Journal of Pure Science. Vol. 15, No. 2. 1-6.

[6] Al-Saphory, R. 2011. Strategic sensors and regional exponential observability. ISRN Applied Mathematics. Vol. 2011. Article ID 673052. pp. 1-13.

[7] Al-Saphory, R., Al-Joubory, M. and Jasim, M. 2013. Regional strategic sensors characterizations. Journal of Mathematical and Computational Science. 3, 401-418.

[8] Boutoulout, A., Bourray, H. and Khazari, A. 2014. Flux observability for hyperbolic systems. Applied Mathematics and Information Sciences Letters. Vol. 2 , No. 1, 13-24.

[9] Brezis, H. 1987. Analyse fonctionnalle; Theorie et applications. 2em tirage. Masson. Paris, France.

[10] Burns, J., Jorggaard, J., Cliff, M. and Zietsman, L. 2009. A PDE approach to optimization and control of high performance buildings, Workshop in Numerical Techniques for Optimization Problems with PDE Constraints, No. 04/2009, DOI: 10.4171/OWR/2009/04. 25-31 January. Houston, Denmark.

[11] Burns, J., Jorggaard, J., Cliff, M. and Zietsman, L. 2010. Optimal sensor design for estimation and optimization of PDE Systems, 2010 American Control Conference. Marriott Waterfront, Baltimore, MD, USA. June 30-July 02, 2010.

[12] Chen Y., Moore K., and Song, Z. 2004. Diffusion boundary determination and zone control via, mobile actuators-sensors networks (MAS-net)-challenge and opportunities. SCOIS. Utah state university, Logan, UT84322, USA.

[13] Curtain, R. F. and Pritchard, A. J. 1978. Infinite dimensional linear theory systems; Lecture Note in Control and Information Sciences. Spring – Verlag, New York.

[14] Curtain, R. F. and Zwart, H. 1995. An introduction to infinite dimensional linear system theory. Springer-Verlag, New York.

[15] El Jai, A. and Pritchard, A. j. 1987. Sensors and actuators in distributed system. International Journal of Control. Vol. 46, 1139-1153.

[16] El Jai, A. and Amouroux, M. 1988. Sensors and observers in distributed parameter systems. International Journal of Control. Vol. 47, 333-347.

[17] El Jai, A. 1991. Distributed systems analysis via sensors and actuators, International Journal of Sensors and Actuators.Vol. 29, 1-11.

[18] El Jai, A.. and Pritchard A. j. 1991. Sensors and controls in analysis of distributed system, Ellis Harwood series in Mathematics and Applications, Wiley, New York.

[19] El Jai, A. and El Yacoubi, S. 1993. On the number of actuators in a parabolic system, Applied Mathematics and Computer Science. Vol.3, No.4, 139-150.

[20] El Jai, A., Simon, M.C. and Zerrik, E. 1993. Regional observability and sensor structures, International Journal of Sensor and Actuator. Vol. 39, No. 2, 95-102.

[21] El Jai, A., Amouroux, M. and Zerrik, E. 1994. Regional observability of distributed syste. International Journal of Systems Science. 1994, Vol. 25, No.2, 301-313.

[22] El Jai, A., Simon, M.C., Zerrik, E. and Amouroux, M. 1995. Regional observability of a thermal proces. IEEE Transaction on Automatic Control. Vol. 40, 518-521.

[23] El Jai A. and Hamzauoi, H. 2009, Regional observation and sensors. International Journal of Applied Mathematics and Computer Sciences. Vol. 19, No. 1, 5-14.

[24] Gressang, R. and Lamont, G.1975. Observers for systems characterized by semi-group. IEEE on Automatic and Control. 20, 523-528.

[25] Hamzauoi, H., 2008, On choice of actuators-sensors structures in the systems, Ph. D. Thesis, Perpignan, France.

[26] Zerrik, E. 1993. Regional analysis of distributed parameter systems, Ph.D. Thesis. University of Rabat, Morocco.

[27] Zerrik, E., Bourray H. and Badraoui L. 2000. How to reconstruct a gradient for parabolic systems. Conference of MTNS 2000. Perignan, France, June-2000, 19-23.

[28] Zerrik, E. and Bourray, H. 2003. Gradient observability for diffusion systems. International Journal of Applied Mathematics and Computer Sciences. Vol. 2, 139-150.

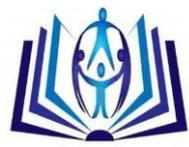

**Author's Biographies**:

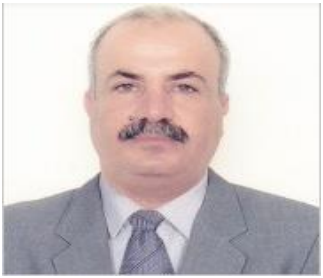

Raheam Al-Saphory is professor associated at the TIKRIT University in IRAQ. He received his Ph.D. degree in Control System and Analysis in (2001) from LTS/ Perpignan University France. He has a post doctorate as a researcher in 2001-2002 at LTS. Al-Saphory wrote one book and many papers in the area of systems analysis and control. Also he is a supervisor of many Ph D. and Msc. students and he was Ex-head of Departement of Mathematics /College of Eduction for Pure Science Tikrit University 2010-2011. He visite many centres and Scientific Departments of Bangor University/ Wales/ UK with academic staff of Iraqi Universities in 2013.

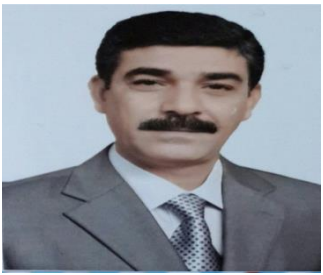

Naseif Al-Jawari is assistant Proffesor at the Al-Mustansiriyah University, IRAQ. He received his Ph.D. degree in Optimal Control Theory in (2000) from Faculty of Mathematics/ Łódź University Poland. Al-Jawari wrote many papers in the area of systems analysis and Optimization. Also he is a supervisor of many Ph D. and Msc. students. He is head of Applied Mathematics Brach/ Department of Mathematics/ College of Science/ Al-Mustansiriyah University 2012-Present.

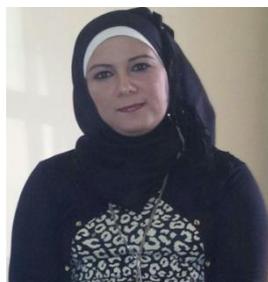

Assma Al-Janabi is a researcher at the Department of Mathematics/ College Science/ Al-Mustansiriyah University/ IRAQ 2012-Present. Here research area focused on Distributed Parameter Systems Analysis and Control. She is obtained here Ms.c. degree from Al-Nahrain University in (2003).